\documentclass[12pt,preprint]{aastex}

\usepackage{natbib}
\usepackage{graphicx}
\usepackage{multirow}
\usepackage{txfonts}

\shorttitle{Estimation of central mass of NGC 4151 with TCAF}
\shortauthors{P. Nandi, S.K. Chakrabarti and S. Mondal}
\begin{document}

\title{Spectral properties of NGC 4151 and the Estimation of black hole mass using TCAF solution}
\author{Prantik Nandi\altaffilmark{1}, Sandip K. Chakrabarti\altaffilmark{1,2}, Santanu Mondal\altaffilmark{3,2}}
\altaffiltext{1}{S. N. Bose National Center for Basic Sciences, JD-Block, Salt Lake, Kolkata, 700098, India}
\altaffiltext{2}{Indian Centre For Space Physics, 43 Chalantika, Garia Station Road, Kolkata, 700084, India}
\altaffiltext{3}{Physics Department, Ben-Gurion University of the Negev, Be'er-Sheva 84105, Israel}

\email{prantiknandi@bose.res.in; sandip@csp.res.in; santanu@post.bgu.ac.il}

\begin{abstract}
We present X-ray spectral analysis of Seyfert 1.5 Active Galactic Nuclei (AGN) NGC~4151 using \textit{NuSTAR} observation during 2012.
This is the first attempt to fit an AGN data using 
the physical Two Component Advective flow (TCAF) solution. 
We disentangle the continuum emission properties of the source in the energy range $3.0$ to $70.0$~keV using 
the spectrum obtained from TCAF model. This model was used as an additive local 
model directly in {\fontfamily{qcr}\selectfont XSPEC}. Additionally, we used a power law (PL) component, 
to take care of possible X-ray contribution from the jet, which is
not incorporated in the present version of TCAF. Our primary aim is to obtain the flow properties and
the mass of the central supermassive black hole from the available archival data. 
 Our best estimate of the average mass obtained from spectral fits of three observations, 
is $M_{BH}=3.03^{+0.26}_{-0.26}\times 10^7 M_\odot$. This is consistent with earlier estimations in the literature such as reverberation mapping, 
gas kinematics and stellar dynamics around black holes. We also discuss the accretion dynamics 
and the flow geometry on the basis of model fitted physical parameters. 
Model fitted disk accretion rate is found to be lower than the low angular momentum halo accretion rate, 
indicating that the source was in a hard state during the observation. 
\end{abstract}


\section{Introduction}
Active Galactic Nuclei (AGNs) exist in a few percent of all massive galaxies in our local universe. 
These AGNs are powered by accretion of matter onto a supermassive black hole (SMBH) 
(mass $\geq 10^6 M_\odot$), generating in the process a huge amount 
of radiation spanning across the electromagnetic spectrum. It is well established that they exhibit 
extreme X-ray flux variability in timescales of minutes to years (Nandra 2001; Turner \& Miller 2009). 
Seyfert galaxies are the sub-classes of AGNs, characterized in terms of total luminosity emitted
from their nuclei. Depending on the characteristics of emission 
lines, AGNs are classified into Type~1 and Type~2 (Osterbrock \& Pogge, 1985). 
The broadband spectral energy distribution (SED) of AGNs is shaped by 
the black hole properties as well as the nature of the infalling matter. 
In the optical-UV band, the SED comprises a `big blue bump', which is broadly consistent with an optically thick 
and geometrically thin accretion disk proposed by Shakura and Sunyaev (1973). In X-ray band, 
AGN consists of a primary continuum, which is approximated by a cut-off PL with a reflection component. 
In general, the spectrum also shows a significant photo-electric absorption, Fe fluorescence (George \& Fabian 1991; Matt et al. 1991) 
and inverse Compton scattering by hot, optically thin Compton cloud (e.g., Haardt \& Maraschi, 1991). 
Along with these components, a cold and/or warm absorber can influence the AGN's X-ray spectrum and an additional soft excess is 
observed in most cases. 
 The origin of this soft X-ray excess bump (Halpern 1984; Singh et al. 1985; Arnaud et al. 1985) is poorly understood in a conventional model. It is believed to be due to further X-ray reflection 
off the accretion disk (Crummy et al. 2006), partially ionized absorption (Gierli\'{n}ski \& Done 2004), or warm Compton upscattering within the accretion disk 
(Magdziarz et al. 1998; Porquet et al. 2004; Schurch \& Done 2006; Dewangan et al. 2007). Few recent works also discussed about the 
warm absorption analyzing different objects data (See, Laha et al. 2016 for details). Sometimes, X-rays coming from the central part of AGNs may be absorbed by a cloud. 
Thus, the soft part of X-ray cannot be seen directly, leading to a classification of less active Type-2 Seyfert AGNs and 
the rest AGNs are classified into Type-1 Seyfert AGNs. However, this cloud is generally transparent in X-rays at energies higher than 4 keV.
It is believed that an Active Galactic Nucleus contains a standard disk component which produces
soft photons and a hot Compton cloud, which upscatters soft photons from the disk and produces a power-law (PL) tail.
So far, many of the models focus on the study of variability and time series analysis even 
though the actual physical origin of variability, change in optical depths, origin and sustenance of the Compton cloud etc. still remained unclear. 

In Chakrabarti (1995), a Two component Advective Flow (TCAF) model based on viscous transonic flow around a black hole was proposed to explain the AGN spectra. Subsequently, the spectral properties in the context of stellar mass black holes was studied in Chakrabarti \& Titarchuk (1995) and Chakrabarti (1997). Out of the two components, one is geometrically thick hot sub-Keplerian component and the other one is the standard Keplerian halo component which is truncated at the inner edge at the centrifugal barrier of the halo. In Fig.~1, we show the cartoon diagram of TCAF model, where soft photons from the Keplerian disk is upscattered by the hot electron cloud formed inside the centrifugal barrier of halo.
According to the TCAF solution, the boundary of the Compton cloud 
is basically the location of the shock (Chakrabarti, 1989) formed in the low-viscosity, sub-Keplerian component.
The post-shock region is called the CENtrifugal barrier supported BOundary Layer or CENBOL which actually dissipates its 
thermal energy through inverse Comptonization of the soft photons intercepted from the Keplerian disk and also supplies matter to outflows and jets.
The steady flow of low-angular momentum matter sustains the Compton cloud or CENBOL. The outer edge of the CENBOL is
also the inner edge of the truncated standard disk. Numerical simulations by Giri \& Chakrabarti (2013) show that 
when the injected flow is sub-Keplerian, as is very likely in case of AGNs, and the viscosity is enhanced on the equatorial plane, the matter segregates into
two components as envisaged by Chakrabarti (1995). Discussions of satisfactory fits of AGN spectra with TCAF are present in the literature (Mandal \& Chakrabarti 2008). This two-component nature of the accretion disk can be treated as a general model where the accretion rates of both the components can be varied independently and the model should be applicable for all the black hole candidates from the usual quasars and AGNs to nano-quasars or stellar-mass black holes. 
The self-consistent fitting of data with TCAF spectra allows one to obtain four flow parameters (accretion rates of the two components, size of the Compton cloud, and the compression ratio of the flow at the shock which determines the optical depth of the cloud) along with the mass of the central object. Recently, a series of papers have been published (Debnath et al. 2014, 2015; Mondal et al. 2014, 2016; Jana et al. 2016; Molla et al. 2016; Chatterjee et al. 2016; Bhattacharjee et al. 2017 and others), which successfully fitted observed data of stellar mass black holes taken from different satellites and described how the flow parameters evolved during outbursts. These authors also obtained the mass of the respective central objects from spectral fits.

The Seyfert 1.5 galaxy NGC~4151 (z=0.00332, de Voucouleurs et al. 1991), sometimes considered as a Seyfert~1 type AGN, is one of the most popular sources for which many AGN phenomena were first characterized (for more details, see Ulrich, 2000). 
It has a bolometric luminosity of $L_{bol}\sim 5\times 10^{43}$ $erg/s$ (Woo \& Urry, 2002). As NGC~4151 is a nearby AGN, 
the mass of the central black hole was estimated by various groups. For example, from the detailed study of C~IV line profile, 
Clavel et al. (1987) found the mass of the central black hole to be $3.7\pm 0.5 \times 10^7 M_\odot$. 
Hicks \& Malkan (2008) measured the dynamical mass $M_{BH}=3.0^{+0.75}_{-2.2} \times 10^7 M_\odot$ from the kinematics of the 
gas in the vicinity of the black hole. Onken et al. (2014) measured the mass of the black hole to be 
$M_{BH}=3.76^{+1.15}_{-1.15}\times 10^7 M_\odot$ by using stellar dynamical mass measurement. Also, many groups 
measured the mass from reverberation mapping and found $M_{BH}=3.57^{+0.45}_{-0.37} \times 10^7 M_\odot$ 
(based on the revised estimate of Grier et al. 2013). 
It is established that the X-ray spectrum of NGC~4151 consists of several components: (1) in medium energy band of X-ray 
(5 - 50 keV), the dominating component in the spectrum is PL. Sometimes, this PL component is extended up to 100~keV. 
It has a photon index in the range of $\Gamma \sim 1.2 - 1.9$ (Ives et al. 1976; Perola et al. 1986; 
Yaqoob \& Warwick 1991; Zdziarski et al. 1996; Beckmann et al. 2005). The other components which are 
observed in this energy range are: a reflection hump in 10-30~keV, a fluorescence Fe~$K_\alpha$ at 6.4~keV 
(Cackett et al. 2014) and the iron edge, (2) In the hard X-ray range ($>$ 50~keV), the PL component steepens with a cut 
off at around 80-100~keV and, (3) in the lower energy range (below 5~keV), the spectrum of NGC~4151 is very complex. 
The soft excess below 2.0~keV is not well understood till date. Holt et al. (1980) took an attempt to extrapolate it by using 
uniformly absorbed PL which has several components: (i) an unabsorbed, non-time varying PL, which is believed to be scattered 
off an extended hot electron cloud, 
(ii) a time-varying PL component (of the order of 60\%) leaks through 
a patchy very thick absorber ($N_H \sim 10^{23} cm^{-2}$), and (iii) an extended component of thermal emission at 
very soft energy (below 0.3~keV) (Weaver et al. 1994; Morse et al 1995; Warwick et al. 1996).

Lubi\'{n}ski et al. (2010) analysed INTEGRAL data during 2003 to 2009 along with other X-ray observations and 
found that the coronal emission had an approximate constant X-ray spectral index and {\it Compton y-parameter}. Authors 
inferred that the corona had an approximately constant geometry and the X-ray source could be situated at the 
base of the jet very much as the CENBOL in TCAF solution mentioned above. NGC 4151 displayed a non-relativistic jet, observed in radio wavelength (Wilson \& Ulvested 1982; Mundell 
et al. 2003; Ulvested et al. 2005) and it was characterized using optical (Storchi-Bergmann et al. 2009) and X-ray observations (Wang et al. 2011a). 

More recent observation using XMM-Newton data reveals that NGC~4151 has a compact coronal structure surrounding a 
maximally rotating (spin parameter, $a=0.998$, Cackett et al. 2014) black hole. 
Zoghbi et al. (2012) shows from reverberation of Fe~$K_\alpha$ that the broad Fe~$K_\alpha$ emission responds to coronal emission 
originating from a height above the accretion disk. Cackett et al. (2014) showed the height of the X-ray source above 
accretion disk to be $h=7.0^{+2.9}_{-2.6}$ $r_g$ using lamp-post model (Martocchia et al. 2002; Miniutti et al. 2003). 
	
In this paper, we present the spectral analysis of \textit{NuSTAR} of NGC 4151 using TCAF model as a local model in XSPEC. Our motivation stems from the fact that 
TCAF is the most general solution of fundamental equations which govern the relativistic flows around a black hole and has been found to be quite
successful in explaining the spectral and temporal properties of stellar mass black holes. Due to its self-consistency, it is capable of measuring all the flow parameters and the mass of the central black hole even with a single observation without any reference to other observations. 
Though TCAF was historically introduced in the context
of AGNs (Chakrabarti, 1995), it has not been applied to analyse their data using XSPEC, though manually data was fitted for M87 (e.g., Mandal \& Chakrabarti, 2008).
This paper will be the first where AGN data would be fitted after TCAF is implemented in XSPEC and an attempt to estimate the 
mass from the spectral data would be made. The paper is organized as follows:
in the next Section, we summarize the observations and data reduction methods and model implementation in XSPEC. In Sec.~3, we 
discuss about the model fitted parameters, their variation during the observing time and estimation of mass. Finally, we 
briefly draw our conclusions. 
		
\section{Data reduction and model fitting}

We analyse \textit{NuSTAR} observational data which are publicly available in {\fontfamily{qcr}\selectfont HEASARC} 
archive with exposure times longer than 5~ksec for NGC~4151 with observation date from November 11 to 14, 2012. 
The observation IDs are 60001111002, 60001111003 and 60001111005 respectively. The details about the observations are 
given in Table~1. 
The same data was analyzed by Beuchert et al. (2017) along with the observation from \textit{Suzaku} and \textit{XMM-Newton} satellites. From their spectral fitting, presence of different absorbing media with the neutral, ionized and highly-ionized absorber were found. Based on different absorption structures authors also found variability of different scales ranging from day to year using their model. 
Keck et al. (2015) used IDR model to fit the \textit{NuSTAR} data along with \textit{Suzaku} data and predicted jet structure for this source. 
They also derived the spin of the black hole $a > 0.9$. From their fit they found that the photon index was varying from 1.25 to 1.74.
\par We analyse \textit{NuSTAR} focal plane module A and module B (FPMA \& FPMB) data using \textit{NuSTAR} data analysis software 
{\fontfamily{qcr}\selectfont NUSTARDAS} (Harrison et al. 2013) version 1.8.0, and {\fontfamily{qcr}\selectfont CALDB} is taken 
from 2018 January 1. The calibration, cleaning and screening of data are performed using standard \textit{nupipeline} script. 
The details about the data extraction procedure is already discussed in Mondal et al. (2016) for the stellar mass
black holes and we follow the same procedure in this paper. We discuss them briefly here for the sake of completeness.
Each of the lightcurves and the spectra was generated from the circular region centered at the source using \textit{nuproducts} script. 
We extracted the source data products for a $40''$ radius circular region centered on NGC~4151. A similar extraction of background 
with a $60''$ radius circular regions on the same detector was chosen to avoid contamination and detector edges. We binned the 
background-subtracted count rates for FPMA and FPMB for 300~sec interval to produce the final light curves. As the detector 
FPMA and FPMB are nearly identical, we use only FPMA's data for further fitting. The response files (\textit{rmf} and 
\textit{arf} files) are generated by using \textit{numkrmf} and \textit{numkarf} modules respectively using in built 
\textit{nuproduct} script. The source and background spectra provide a source dominated view for 3 - 70~keV energy for NGC~4151. It  
is shown in Fig.~2. We reprocessed all data using {\fontfamily{qcr}\selectfont HEAsoft} version 6.22.1 (Arnaud 1996), 
which includes {\fontfamily{qcr}\selectfont XSPEC v12.9.1p}.

For the spectral analysis, we use {\fontfamily{qcr}\selectfont XSPEC} spectral fitting software package with standard cosmological 
parameters, e.g., Hubble constant ($H_0=$) 70~km~sec$^{-1}$Mpc$^{-1}$, Matter density parameter, $(\Omega_M=)$ $0.27$ and Dark energy density parameter $(\Omega_\Lambda=)$ $0.73$ (Komatsu et al. 2011). We also use the X-ray cross-section value from Verner et al. (1996) and the Milky Way absorption column density to the weighted-average value for NGC 4151 $N_H=2.3\times10^{22}$ atoms cm$^{-2}$ from the LAB survey (Kalberla et al. 2005).
For the model fitting, we use the spectra coming from TCAF model as a basic model along with an additional PL component, to take care of possible contribution from the base of the jet which is not included in TCAF spectrum code. To fit the data  
we need a Gaussian component at 6.4~keV energy. Keck et al. (2015) fitted the spectrum of NGC 4151 with a 
blended Fe~XXV and Fe~XXVI K$_\beta$ ($E=7.88$ keV and $E=8.25$ keV respectively) absorption lines. For this, we use $gabs$, 
which has a peak at $\sim$ 8.2~keV. Thus our composite model to fit the spectrum is $wabs\*(TCAF+Gaussian+PL)\* gabs$. 
It is reported that the hydrogen column density $(N_H)$ along the line of sight of NGC 4151 rapidly varies with time 
ranging from $1.4\times10^{22}$ to $10.2\times10^{22}$~atoms~cm$^{-2}$ (Puccetti et al. 2007).   
During the spectral fit with TCAF, we supply five model parameters, viz., (i) black hole mass in unit of solar mass ($M_\odot$) unit, 
(ii) Keplerian disk accretion rate ($\dot{m}_d$) in unit of Eddington rate ($\dot{M}_{EDD}$), (iii) Sub-Keplerian halo accretion 
rate ($\dot{m}_h$) in unit of Eddington rate ($\dot{M}_{EDD}$), (iv) shock compression ratio (R), and (v) shock location ($X_s$) 
in unit of Schwarzschild radius ($r_g=2GM/c^2$). We put all the parameters in a data file (lmodel.dat) as an input to run the source 
code using \textit{initpackage} and \textit{lmod} tasks in XSPEC. We ran the TCAF code for a vast number of input parameters
for fitting purpose and generated many spectra from which best fit was obtained (see, Chakrabarti, 1997 for examples
of spectral variations with flow parameters).

\section{Mass estimation by TCAF model}	

We now discuss the results based on spectral fitting with TCAF. Before that, we briefly present the 
principle which TCAF follows to fit the spectra of any black hole candidate from a single data and then we discuss
how the mass is computed from the fit. TCAF is a special case of viscous transonic flow solution originally proposed by Chakrabarti (1990) in which
it was shown that depending on viscosity parameter, a flow may become a Keplerian disk or remain advective and form centrifugal pressure driven
shock wave close to a black hole. Since turbulent viscosity is higher on the equatorial plane, the Keplerian disk will form on the
equator and is surrounded by the advective component. However, TCAF does not require viscosity explicitly, since it assumes that the
job of viscosity is to segregate two rates from one advective flow. Thus instead of using one rate and one viscosity, it uses 
two separate accretion rates for the Keplerian and the sub-Keplerian components. Chakrabarti (1995) proposed that this two component disk 
could be useful explaining AGN spectra. The spectra of TCAF were computed self-consistently in Chakrabarti and Titarchuk (1995) and Chakrabarti (1997)
where it was shown that only four flow parameters, namely the accretion rates of the two accretion components, 
shock location (describing the size of the Compton cloud), and shock strength (deciding the optical depth in conjunction with the accretion rates
and shock location) are enough to get a spectrum if the mass is known. In obtaining a spectrum, we use the radiation from the Keplerian component as the seed photons. The shocks in 
vertical equilibrium gives the amount of the seed photons intercepted by the post-shock region. The Comptonization of
the seed photons were self-consistently computed from formalisms given in Sunyaev and Titarchuk (1980). Since the number density of seed photons,
electrons in the post-shock region (acting as the Compton cloud), shock location, etc. all depend on the mass of the black hole,
TCAF spectral fit directly depends on the mass of the central black hole. Thus, in the absence of any knowledge of the mass,
TCAF is capable of fitting the spectrum and obtain the black hole mass along with other flow variables mentioned above from each observation. 

Occasionally, a data may have additional sources of radiations not included in TCAF. In that case we use a PL component in addition to TCAF generated spectrum
to take care of the X-rays contribution from the jets, 
since TCAF does not include a jet in the present version. This increases the number of parameters
from four to six, when the mass is known.
In Fig.~3(a-c), we show the TCAF model fitted $3.0-70.0$~keV spectra along with residual in the bottom panel for three observation IDs. 
In Fig.~3d, unabsorbed model spectra are shown which are used to fit the observed data. The corresponding model fitted parameters are given in Table~2.  

The model fitted parameters give a physical understanding about accretion flow dynamics and radiation processes 
around the central black hole of NGC~4151. The variations of different parameters with day are shown 
in Fig.~4. Panel (a) shows the model fitted mass of the central black hole, independently obtained from each fit. This is more or 
less constant for all of the three IDs and the average mass becomes $3.0^{+0.2}_{-0.2}\times10^7 M_\odot$. 
Procedure of mass estimation is given below.
Next two panels (b) and (c) show variations of accretion rates (disk and halo accretion rates). One can see that for all the
observations, the disk rate is always very low as compared to the halo rate. This is an indication that the 
electron number in CENBOL (acting as the Compton cloud in TCAF) is too high and they could not be cooled by the
intercepted soft photons coming from the Keplerian disk. This is why the object was always in the hard state. The spectral state 
is also confirmed from the variation of other three parameters such as $R$, $X_s$ and $\alpha$  shown respectively in panels (d) (e) and (g). 
The value of $X_s$ is more or less constant which agrees with the earlier findings of constant corona (Lubi\'{n}ski et al. 2010).
The panel (f) shows the variation of TCAF normalization which is also nearly constant, as expected from TCAF solution,
as it is a scale factor between the photons emitted in the rest frame of the object as those observed by an instrument
on earth, provided absorption is taken care of. A similar method has been used to obtain masses of several black holes in X-ray binary systems  
(Jana et al. 2016; Chatterjee et al. 2016; Molla et al. 2017; Bhattacharjee et al. 2017). 
In the last panel of Fig.~4, we show the variation of PL normalization.nIt is to be noted that there are other models 
in the literature which estimate the mass of the black holes: In the so-called scaling model by
Titarchuk \& Fiorito (2004), one requires to have numerous data points at various intensity
level of the source so as to identify, the 
presence of a break frequency and the PL-index saturation. It also requires a reference
mass. Due to paucity of data in the present source, and the fact that for supermassive black holes, changes in spectral state and PL index might take very long time, this method
could not be used to compare the mass. In the case of stellar mass black holes with profuse observed data with satisfactory break frequency and PL saturation index, such a scaling mathod gives a compatible result as TCAF (Molla et al. 2017).

 A distinct difference between the fit with TCAF and other models is that, since TCAF 
already includes mass in obtaining the entire spectrum, the whole 
spectrum is obtained with a constant normalization across the spectral states. 
The reflection component is also included in a TCAF spectrum since the exact number 
of photons participating in reflections is known from CENBOL size. Other models
require separate normalization for each additive components required to obtain the 
whole spectrum. The constant normalization constraint
enables us to obtain the central mass. This is done in the following way: 
First, we fit all three spectra using TCAF by keeping all parameters free. 
We obtained the TCAF normalization in a narrow range: $0.0545$ to $0.0581$. This means the masses obtained are accurate.
As the mass is an independent parameter, we obtained the mass in the range $2.99\times 10^7$ to $3.06 \times 10^7$ $M_\odot$. 
After this, we also consider a constant normalization of $0.0570$, which is the average of value coming from all the IDs. 
Again we fit the data using this average normalization and we obtained the black hole mass from $2.99 \times 10^7-3.07 \times 10^7$ $M_\odot$.
 To further verify the correctness of the estimated mass, we consider a mass range from 
$1.0\times 10^4 M_\odot$ to $2.0 \times 10^8 M_\odot$ and refit all the spectra by freezing all model 
parameters as constant to observe the variation of reduced $\chi^2$ ($\chi^2_{red}$). To consider the best fit, we restricted $\chi_{red}^2$ at $\chi^2_{red}\leq 1.5$. The variations of $\chi^2_{red}$ for all the spectra is shown in Fig.~5(a-c). 
In panel (d), we show the comparison of $\chi_{red}^2$ variation with mass for three separate observations. From all spectral fits, it is clear that the minima are at $\sim 3.0^{+0.2}_{-0.2} \times 10^7 M_\odot$.

\section{Discussions and concluding remarks}

In this paper, we study spectral properties of NGC~4151 with TCAF using three \textit{NuSTAR} observations during 
2012. For the first time, we use TCAF model directly in XSPEC to fit the AGN data. We obtain the accretion flow 
properties as extracted by TCAF. For the best fit, we also use an additional PL component to consider the
possible contribution from the jet, which is not currently incorporated in TCAF.
The average of the estimated mass of the central black hole appears to be $3.03^{+0.26}_{-0.26}\times 10^7 M_\odot$ 
which is in the same ball park as obtained by previous workers from other considerations. 
However, unlike other models, we obtained the accretion rates
in the Keplerian and the halo components and found that the halo rate is much larger as compared to the disk accretion rate,
consistent with the fact that the object was always in the hard state and the fact that Active galactic 
nuclei usually accretes low angular momentum matter, unless a part is also converted to 
Keplerian by a significant viscosity. We also observe that in all the 
three observations, the shock location does not really change. In other words, the so-called Compton cloud 
(or, `corona' in some models) remained steady. 
This is partly due to the fact that in a supermassive black hole environment, quick variations are not possible.
However to get the more detail understanding about the evolution of the accretion flow, 
we need to have a long term monitoring. 

  Of course, our estimations of the flow parameters are valid if a TCAF is formed, i.e., the
 advective flow develops a centrifugal barrier close to the black hole. The assumption is valid if much of the accretion matter is not Keplerian, and has very low angular momentum.
 Judging from the goodness of fit, we believe the flow is indeed having TCAF configuration 
and the halo is dominant. Since phenomenological models also use a Compton cloud and soft photon source 
in one way or the other in order to fit the data, they may also achieve good fits. However, flow properties 
will come only if the radiative properties are coupled to hydrodynamics as is done when TCAF is used fitting.

It is reported that NGC~4151 has a highly spinning black hole (Cackett et al. 2014) at the center. 
From the fitting, we see that the shock location (Compton cloud size) required for a good fit 
is  $\sim 150~r_s$, which is far away from the black hole, as expected from a hard state.
Thus the spin is not likely to change our conclusions regarding the flow parameters 
and the mass. However, spin effects should be taken into account
when the object approaches a softer state and particularly 
if a double horned iron lines are required to  fit the data and the flow is relativistic.

\section*{Acknowledgments}
PN acknowledges CSIR fellowship for this work. SM acknowledges funding from the IAEC-UPBC joint research foundation (grant No. 300/18), and support by the Israel Science Foundation (grant No. 1769/15) and the hospitality of S. N. Bose Centre during his visit. 
SM is also thankful to Keith A. Arnaud for helping in model inclusion during his visit to NASA/GSFC as a student of 
COSPAR Capacity-building Workshop Fellowship program jointly with ISRO.
This research has made use of the NuSTAR Data Analysis Software (NuSTARDAS) jointly developed by the ASI Science Data Center (ASDC, Italy) and the California Institute of Technology (Caltech, USA).
{}

\begin{figure}
\vskip 0.5cm	
\begin{center}
\includegraphics[height=3.5cm, width=8.0cm, angle=0]{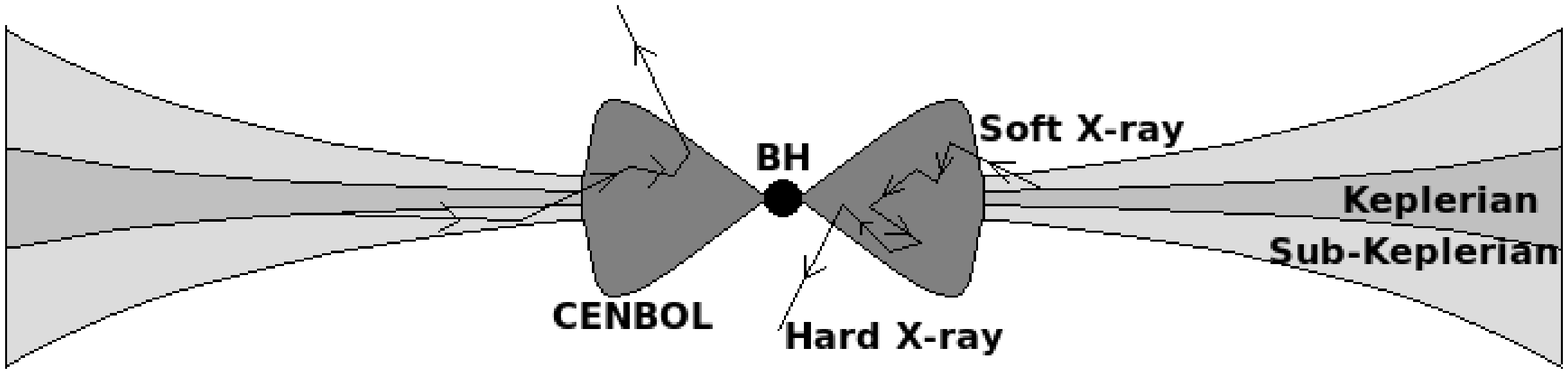}
\caption{Cartoon diagram of TCAF model (CT95), where Keplerian disk is at the equatorial plane producing soft photons, 
	which are upscattered by the CENBOL formed by sub-Keplerian hot flow.
}

\end{center}
\end{figure}

\begin{figure}
\vskip 0.5cm	
\begin{center}
\includegraphics[height=5.5cm, width=8.0cm, angle=0]{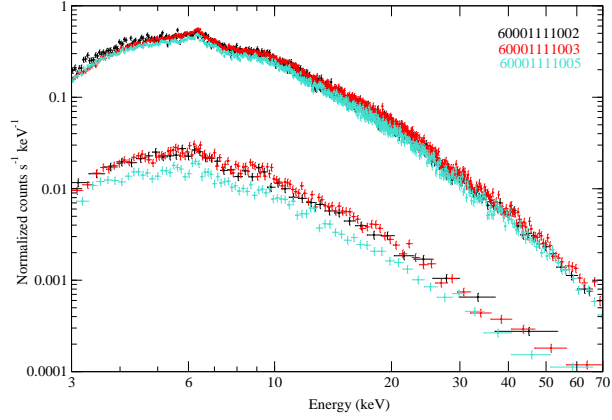}
\caption{The source and background spectra are plotted with error bars for the three data IDs. 60001111002 (black), 60001111003 (red) and 60001111005 (turquoise).}
\end{center}
\end{figure}

\begin{figure}
\vskip 0.5cm
\begin{center}
\includegraphics[height=5.5cm, width=8.0cm, angle=0]{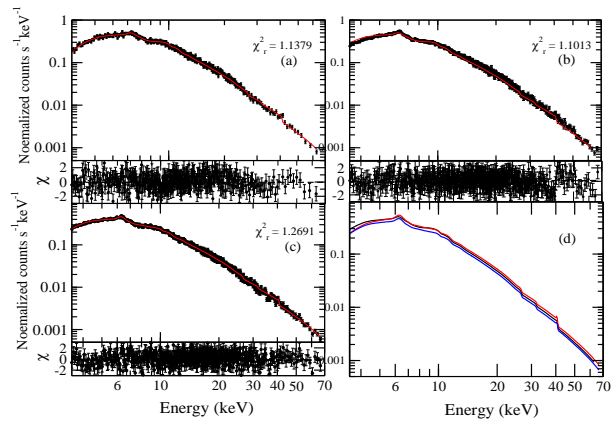}
\caption{TCAF model fitted 3.0 - 70~keV \textit{NuSTAR} spectra with variation of $\chi^2$, for three IDs. (a) 60001111002, 
(b) 60001111003 and (c) 60001111005. In (d), TCAF+PL model generated spectra are shown, which are used to fit the observation.}
\end{center}
\end{figure}

\begin{figure}
\vskip 0.5cm	
\begin{center}
\includegraphics[height=5.5cm, width=8.0cm, angle=0]{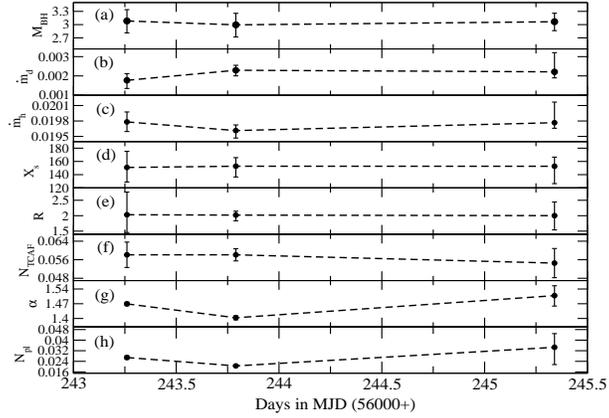}
\caption{Variation of model fitted parameters (a) mass of the black hole in unit of $10^7 M_\odot$, (b) and (c) are the disc rate 
and halo rate respectively. In (d) shock location, (e) shock compression ratio, (f) TCAF model normalization, (g) photon index,
and (i) PL model normalization.}
\end{center}
\end{figure}

\begin{figure}
\begin{center}
\includegraphics[height=5.5cm, width=8.0cm, angle=0]{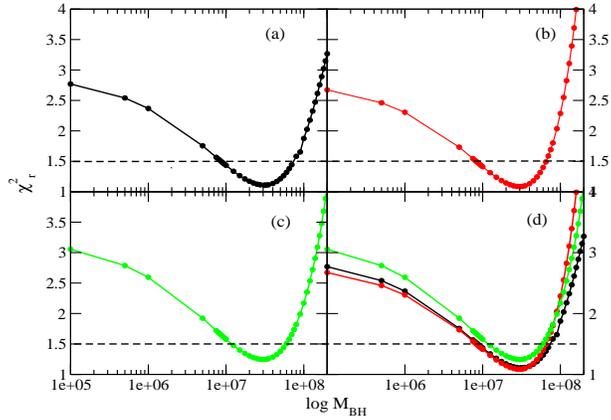}
\caption{In (a-c), variation of $\chi_r^2$ with black hole mass in unit of solar mass ($M_\odot$) for three different observation IDs. 
In(d), shows the comparison of variation for three IDs.}
\end{center}
\end{figure}

\begin{table}
	\caption{\label{table1} Observation log of \textit{NuSTAR} data}
	\vskip -0.2 cm
	\begin{center}
		\begin{tabular}{c c c c c c c c c}
			\hline
			$ \rm Date $ & $\rm Time (UT) $ & $ \rm Observations $ & $ \rm Exposure $ & $\rm Count Rate$\\
			$(yyyy-mm-dd)$ & $(hh-mm-ss)$ & $\rm ID.$ & $\rm Time (sec)$ & $\rm (cts/s)$\\
			\hline
			$ $ & $ $ & $  $ & $  $ & $ $\\
			$\rm 2012-11-12$ & $\rm 06:06:07$ & $ \rm 60001111002$ & $\rm 21864 $ & $\rm 4.259 \pm 0.014$\\
			$ $ & $ $ & $  $ & $  $ & $ $\\
			$\rm 2012-11-12$ & $\rm 18:56:07$ & $ \rm 60001111003$ & $\rm 57036 $ & $\rm 4.309 \pm 0.014$\\
			$ $ & $ $ & $  $ & $  $ & $ $\\
			$\rm 2012-11-14$ & $\rm 08:16:07$ & $ \rm 60001111005$ & $\rm 61531 $ & $\rm 3.768 \pm 0.008$\\
			$ $ & $ $ & $  $ & $  $ & $ $\\
			\hline
		\end{tabular}
	\end{center}
\end{table}

\begin{table}
	\caption{\label{table2} TCAF\textbf{+PL} Model Fitted Parameters in 3.0-70.0~keV energy band for NGC 4151.}	
	\vskip -0.2 cm
	{\centerline{}}
		\scriptsize
	\begin{center}
		\begin{tabular}{c c c c c c c c c c}
			\hline
			$\rm ID. $& $\rm \dot{m}_d $ & $ \rm \dot{m}_h $ & $ X_s $ & $ R $ & $ M_{BH} $ & $ N_{TCAF} $ & $ \alpha $ & $ N_{pl} $ & $\chi^2/dof$\\
			$ $ & $\rm (10^{-3}) $ & $ \rm (10^{-2}) $ & $ (r_g) $ & $  $ & $ \rm (10^{7}M_\odot) $ & $ (10^{-2}) $ & $  $ & $ (10^{-2}) $ & \\
			\hline
			$ $& $ $&$ $&$ $&$ $&$ $&$ $&$ $&$ $& \\
			$\rm60001111002  $& $\rm1.77^{+0.40}_{-0.34}  $ & $ \rm 1.98^{+0.02}_{-0.02} $ & $ 150.87^{+22.0}_{-24.4} $ & $ 2.03^{+0.58}_{-0.73} $ & $ 3.06^{+0.25}_{-0.27} $ & $ 5.81^{+0.55}_{-0.54} $ & $ 1.47^{+0.01}_{-0.01}$ & $ 2.69^{+0.13}_{-0.13} $ & $471.11/414$\\
			$ $& $ $&$ $&$ $&$ $&$ $&$ $&$ $&$ $& \\
			$\rm60001111003 $& $\rm2.29^{+0.29}_{-0.26}  $ & $ \rm 1.96^{+0.01}_{-0.01} $ & $ 152.73^{+13.0}_{-16.2} $ & $ 2.01^{+0.12}_{-0.18} $ & $ 2.99^{+0.27}_{-0.26} $ & $ 5.83^{+0.26}_{-0.26} $ & $ 1.40^{+0.01}_{-0.01}$ & $ 2.05^{+0.50}_{-0.51} $ & $681.10/633$\\
			$ $& $ $&$ $&$ $&$ $&$ $&$ $&$ $&$ $& \\
			$\rm60001111005 $& $\rm2.21^{+0.31}_{-0.99}  $ & $ \rm 1.97^{+0.01}_{-0.04} $ & $ 152.37^{+26.2}_{-13.8} $ & $ 2.02^{+0.44}_{-0.46} $ & $ 3.06^{+0.27}_{-0.26} $ & $ 5.45^{+0.63}_{-0.63} $ & $ 1.51^{+0.05}_{-0.05}$ & $ 3.46^{+0.10}_{-0.13} $ & $743.38/586$\\
			$ $& $ $&$ $&$ $&$ $&$ $&$ $&$ $&$ $& \\
			\hline
		\end{tabular}
		\newline
		\noindent{Here we list the variation of disk accretion rate ($\dot{m}_d$) and halo accretion rate ($\dot{m}_h$) in Eddington units, shock location ($X_s$) in} 
		\noindent{Schwarzschild radius ($r_g$), shock compression ratio ($R$), mass of the central black hole ($M_{BH}$) in $10^7 M_\odot$,} 
		\noindent{TCAF normalization ($N_{TCAF}$), PL index ($\alpha$), and \textbf{PL} normalization ($N_{pl}$) with IDs along with their error bars.} 
		\noindent{The $\chi^2$ per degrees of freedom ($dof$) for each cases is given in the last column.}
	\end{center}
\end{table}
\begin{table}
	\caption{\label{table3} The TCAF parameter space is defined in the file lmod.dat. }	
	\vskip -0.2 cm
	{\centerline{}}
	\begin{center}
		\begin{tabular}{c c c c c c c c}
			\hline
			\rm Model parameters & \rm Parameter units  &  \rm Default value  & \rm Min.  & Min.  &  Max.  &  Max.  & Increment \\
			\hline
			$\rm M_{BH} $& $\rm M_{Sun} $ & $ \rm 3.6\times10^7 $ &  $2\times10^6$  &  $2\times10^6$  &  $5.5\times10^8$  &  $5.5\times10^8$  & $ 10.0 $ \\
			$\rm \dot{m}_d $& $\rm Edd $ & $ \rm 0.001 $ & $ 0.0001 $ & $ 0.0001 $ & $ 1.0 $ & $ 1.0 $ & $ 0.0001 $ \\
			$\rm \dot{m}_h $& $\rm Edd $ & $ \rm 0.01 $ & $ 0.0001 $ & $ 0.0001 $ & $ 2.0 $ & $ 2.0 $ & $ 0.0001 $ \\
			$\rm X_s $& $\rm r_g $ & $ \rm 200.0 $ & $ 10.0 $ & $ 10.0 $ & $ 500.0 $ & $ 500.0 $ & $ 2.0 $ \\
			$\rm R $& $\rm  $ & $ \rm 1.7 $ & $ 1.1 $ & $ 1.1 $ & $ 6.8 $ & $ 6.8 $ & $ 0.1 $ \\
			
			\hline
			
		\end{tabular}
	\end{center}
\end{table}

\end{document}